\documentclass[superscriptaddress,showpacs,a4paper,twocolumn]{revtex4}
\usepackage{graphicx}
\usepackage[latin1]{inputenc}
\usepackage{multirow}

\begin{document}
\title{Hydrogen molecular ions for improved determination of fundamental constants}

\author{J.-Ph.~Karr}
\author{L.~Hilico}
\affiliation{Laboratoire Kastler Brossel, UPMC-Univ. Paris 6, ENS, CNRS, Coll\`ege de France, 4 place Jussieu, F-75005 Paris, France}
\affiliation{Universit\'e d'Evry-Val d'Essonne, Boulevard Fran\c cois Mitterrand, F-91000 Evry, France}
\author{J.C.J.~Koelemeij}
\affiliation{Department of Physics \& Astronomy and LaserLaB, VU University, De Boelelaan 1081, 1081 HV Amsterdam, The Netherlands}
\author{V.I. Korobov}
\affiliation{Bogoliubov Laboratory of Theoretical Physics, Joint Institute for Nuclear Research, Dubna 141980, Russia}

\date{\today}
\begin{abstract}
The possible use of high-resolution rovibrational spectroscopy of the hydrogen molecular ions H$_2^+$ and HD$^+$ for an independent determination of several fundamental constants is analyzed. While these molecules had been proposed for metrology of nuclear-to-electron mass ratios, we show that they are also sensitive to the radii of the proton and deuteron and to the Rydberg constant at the level of the current discrepancies colloquially known as the proton size puzzle. The required level of accuracy, in the 10$^{-12}$ range, can be reached both by experiments, using Doppler-free two-photon spectroscopy schemes, and by theoretical predictions. It is shown how the measurement of several well-chosen rovibrational transitions may shed new light on the proton-radius puzzle, provide an alternative accurate determination of the Rydberg constant, and yield new values of the proton-to-electron and deuteron-to-proton mass ratios with one order of magnitude higher precision.
\end{abstract}
\maketitle

From Bohr's model of the atom to the advent of quantum electrodynamics (QED), precision spectroscopy of atomic hydrogen has played a key role in our understanding of matter and its interaction with light. Since the first measurement of the Lamb shift in 1947~\cite{lamb1947,bethe1947}, the predictions of QED have been verified with an increasing level of accuracy which, together with stringent tests in other areas of physics, led to assume the validity of this theory and use it to extract the values of fundamental physical constants from experimental data~\cite{codata2010}. Specifically, available data on the hydrogen (H) and deuterium (D) atoms are used to extract the Rydberg constant $R_{\infty}$ and the charge radii of the proton ($r_p$) and deuteron ($r_d$). Data from electron-proton and electron-deuteron scattering experiments also contribute in this determination.

Recently, the undisputed status of these results has been challenged by the measurement of the Lamb shift in muonic hydrogen~\cite{pohl2010,antognini2013}. The very precise value of $r_p$ deduced from this experiment is in strong disagreement with previous determinations. The discrepancy with the CODATA adjustment~\cite{codata2014} amounts to $5.6\sigma$, or to $4.5\sigma$ if only the H and D data are taken into account ~\cite{codata2010}. Similar discrepancies on the deuteron radius can be inferred through the very precise determination of $r_d^2 - r_p^2$ from the 1S-2S H/D isotopic shift~\cite{jentschura2011}. Although many efforts have been undertaken in the last few years, no convincing solution of the ''proton size puzzle'' has been found so far (see~\cite{pohl2013} for a review). One of the possible explanations is that the error bars, both of hydrogen spectroscopy and scattering experiments~\cite{horbatsch2016} were underestimated. New scattering experiments are in preparation or underway, including electron-proton~\cite{gasparian2014}, electron-deuteron~\cite{kohl2015} and muon-proton scattering~\cite{kohl2014}. In atomic hydrogen, the 1S-3S(D)~\cite{galtier2015,yost2016}, 2S-2P~\cite{vutha2015}, and 2S-4P~\cite{beyer2015} transitions are under study in order to cross-check and improve previous results. An independent determination of $R_{\infty}$, which is strongly correlated to $r_p$ and $r_d$, is another way to shed new light on this problem. Experiments on helium atoms and He$^+$ ions~\cite{vanrooij2011,herrmann2009,kandula2011}, as well as highly charged hydrogenlike ions~\cite{tan2011} may ultimately achieve this goal. On a more general level, the proton size puzzle exemplifies how improved and independent determinations of fundamental physical constants from different physical systems provide essential cross checks of our understanding of the physical world. In this work, we propose a new route towards an independent determination of the Rydberg constant, nuclear radii and nucleus-to-electron mass ratios, relying on high-resolution laser spectroscopy of the hydrogen molecular ions (HMI) H$_2^+$ and HD$^+$.

These systems have long been identified as promising for metrology of the proton-to-electron mass ratio $\mu_{pe}$~\cite{wing1976,koelemeij2007}. Recently, the measurement of a one-photon rovibrational transition in HD$^+$ led to a determination of $\mu_{pe}$ with 2.9~ppb~\cite{biesheuvel2016}. The current CODATA uncertainty on $\mu_{pe}$ is indeed the leading source of uncertainty of the theoretical transition frequencies~\cite{korobov2014a,korobov2014b}, making it the correct parameter to constrain from a single measurement. However, similarly to atomic hydrogen, transition frequencies in HMI also depend on $R_{\infty}$ and $r_p$ with, in the HD$^+$ case, additional dependencies on $r_d$ and on the deuteron-to-proton mass ratio $\mu_{dp}$. This means that a combination of $n$ measurements on distinct transitions (with $n=2, 3, 4, 5$) in H$_2^+$ and/or HD$^+$ may allow the determination of up to $n$ constants out of the set of five constants $\{ R_\infty, \mu_{pe}, \mu_{dp}, r_p, r_d \}$.

On the experimental side, the most attractive feature of this approach is that it only relies on Doppler-free frequency measurements of extremely narrow transitions. Indeed, rovibrational states supported by the ground 1s$\sigma$ electronic curve have long lifetimes of the order of days~\cite{peek1979,pilon2012} in H$_2^+$ and tens of milliseconds~\cite{amitay1994,pilon2013} in HD$^+$ . This is a significant advantage with respect to atomic hydrogen, where only the 1S-2S transition has a small natural width and can be measured with the highest accuracy~\cite{parthey2011}, and a second measurement on a much wider transition such as 2S-8S(D) or 2S-12D~\cite{debeauvoir1997,schwob1999}, involving an intricate analysis of systematic effects, is required for a joint determination of $R_{\infty}$ and $r_p$.

The first step is to identify rovibrational transitions suitable for high-resolution spectroscopy. Throughout the paper, rovibrational states are denoted by $(v,L)$, where $v$ and $L$ are the vibrational and rotational quantum numbers, and are assumed to be supported by the ground electronic curve. Up to now, only one-photon transitions in HD$^+$ have been observed~\cite{wing1976,koelemeij2007,bressel2012,biesheuvel2016} with a precision limited to the ppb range by Doppler broadening. To overcome this, our proposal considers only Doppler-free two-photon transitions in ensembles of trapped and sympathetically cooled HMI, which would allow improving the accuracy by several orders of magnitude. The transitions can be detected by resonance-enhanced multiphoton dissociation (REMPD), as was done in previous experiments~\cite{koelemeij2007,bressel2012,biesheuvel2016}.

In the case of H$_2^+$, $(v,L) \to (v' = v+1,L')$ two-photon transitions are the most favorable since the transition strength rapidly decreases with $\Delta v =|v'-v|$, as shown in~\cite{hilico2001}. The other main requirement is efficient preparation of the ions in the initial state of the transition, which can be achieved by resonance-enhanced multiphoton ionization (REMPI) of H$_2$. Highly selective ion production was demonstrated for $0 \leq v \leq 6$ and $L = 1,2$~\cite{ohalloran1987}; we choose $L=2$ as these states have a simpler hyperfine structure (two sublevels as compared to five)~\cite{karr2008}. The seven selected transitions are listed in Table~\ref{sensith2}; spectroscopy of the $(0,2) \to (1,2)$ transition is being pursued at LKB Paris~\cite{karr2008,karr2012}.

The fact that one-photon dipole transitions are weakly allowed opens up different avenues for spectroscopy of HD$^+$. It has been proposed to probe a two-photon transition with quasi-degenerate counterpropagating photons~\cite{tran2013}, where the lasers are tuned close to resonance with an intermediate rovibrational level in order to enhance the transition probability. In this case, the residual first-order Doppler broadening is suppressed by the two-photon Lamb-Dicke effect. State-selected ion production is not required: HD$^+$ ions can be obtained by electron-impact ionization of HD, after which they will relax to $v=0$ within a few hundreds of milliseconds, ensuring sufficient population in the states $(0,L)$ with $L \leq 5$ at 300~K; moreover the REMPD signal is enhanced by the interaction with blackbody radiation, which continuously recycles ions from other rotational states into the desired state~\cite{tran2013, biesheuvel2016}. Four transitions from $v=0$ with an intermediate level  $(v'',L'')$ lying sufficiently close to the midpoint energy $(E_{(v,L)} + E_{(v',L')})/2$ have been identified (see~\cite{karr2005}) and are listed in Table~\ref{sensithd}. An experiment to measure the $(0,3)\rightarrow(4,2)\rightarrow(9,3)$ transition frequency is currently underway at VU University Amsterdam.

\begin{table}
\small
\begin{center}
\begin{tabular}{|c|c|c|c|c|c|}
\hline
Name & $(v,L)$ & $(v',L')$ & $\lambda$ ($\mu$m) & $s_{\mu_{pe}}$ & $10^9 \; s_{r_p}$  \\
\hline
H2(0) & (0,2) & (1,2) & 9.1661 & -0.4657 & -1.240  \\
H2(1) & (1,2) & (2,2) & 9.7321 & -0.4346 & -1.216  \\
H2(2) & (2,2) & (3,2) & 10.350 & -0.4013 & -1.194  \\
H2(3) & (3,2) & (4,2) & 11.031 & -0.3652 & -1.173  \\
H2(4) & (4,2) & (5,2) & 11.787 & -0.3252 & -1.153  \\
H2(5) & (5,2) & (6,2) & 12.636 & -0.2801 & -1.133  \\
H2(6) & (6,2) & (7,2) & 13.603 & -0.2279 & -1.114  \\
\hline
H     & \multicolumn{3}{|c|}{H(1S-2S)} & 0.00054 & -0.8502  \\
\hline
\end{tabular}
\caption{Selected ro-vibrational transitions $(v,L) \to (v',L')$ in H$_2^+$. The lower and upper rovibrational levels and the transition wavelength are given in the first three columns. The relative sensitivities of the transition frequency on $\mu_{pe}$ and $r_p$ (defined by Eq.~(\ref{defsensit})) are given in the last two columns. The sensitivities of the 1S-2S transition in H, obtained from the results compiled in~\cite{codata2010}, are given in the last line. \label{sensith2}}
\end{center}
\end{table}

\begin{table*}
\small
\begin{center}
\begin{tabular}{|c|c|c|c|c|c|c|c|c|c|c|}
\hline
Name & $(v,L)$ & $(v'',L'')$ & $(v',L')$ & $\lambda_1$ ($\mu$m) & $\lambda_2$ ($\mu$m) & $\lambda_{\rm eff}$ ($\mu$m) & $s_{\mu_{pe}}$ & $s_{\mu_{dp}}$ & $10^9 \; s_{r_p}$ & $10^9 \; s_{r_d}$ \\
\hline
HD(R) & (0,0) & (4,1) & (0,2) & 1.4040 & 1.4304 & 76.149 & -0.9848 & -0.3284 & -1.058 & -6.335  \\
HD(2) & (0,1) & (1,0) & (2,1) & 5.3501 & 5.3857 & 809.34 & -0.4601 & -0.1534 & -0.619 & -3.701  \\
HD(4) & (0,5) & (2,4) & (4,5) & 2.8764 & 2.8606 & 518.60 & -0.4179 & -0.1394 & -0.601 & -3.587  \\
HD(9) & (0,3) & (4,2) & (9,3) & 1.4424 & 1.4453 & 730.58 & -0.3522 & -0.1175 & -0.588 & -3.500  \\
\hline
H-D   & \multicolumn{6}{|c|}{D(1S-2S)-H(1S-2S)}          & -0.9992 &  1.0013 & 3125   & -18722  \\
\hline
\end{tabular}
\caption{Selected ro-vibrational transitions in HD$^+$. The lower, intermediate and upper rovibrational levels are given in the first three columns. The next two columns display the wavelengths $\lambda_1,\lambda_2$ of the nondegenerate two-photon transition, and the effective wavelength $\lambda_{\rm eff} = \left| 1/\lambda_1 - 1/\lambda_2 \right|^{-1}$ for absorption of counterpropagating photons. The last four columns give the relative sensitivities of the transition frequency on $\mu_{pe}$, $\mu_{dp}$, $r_p$, and $r_d$ (defined by Eq.~(\ref{defsensit})). Note that the first transition $(0,0) \to (0,2)$ is a stimulated Raman transition with copropagating photons. The sensitivities of the hydrogen-deuterium isotopic shift of the 1S-2S transition, obtained from~\cite{jentschura2011}, are given in the last line. \label{sensithd}}

\end{center}
\end{table*}

The next step is to compute the dependence of the transition frequencies on fundamental constants. The energy of rovibrational levels of HMI, calculated in the framework of QED, may be written as:
\begin{equation}
E = R_{\infty}\!\left[ E_{\rm nr} (\mu_n) \!+ \alpha^2 F_{\rm QED} (\alpha) + \sum_{n} A^{\rm fs}_n (r_n / a_0)^2 \right] \label{energy}
\end{equation}
where $\alpha$ is the fine-structure constant, and $a_0 = \alpha/4\pi R_{\infty}$ is the Bohr radius. The main contribution to $E$ is the nonrelativistic (Schr\"odinger) energy $E_{\rm nr} (\mu_n)$, which depends on the mass ratio(s) $\mu_n = \mu_{pe}$ in H$_2^+$, and $\mu_n = \{ \mu_{pe}, \mu_{dp} \}$ in HD$^+$; the sensitivity coefficients $\partial E_{\rm nr}/\partial \mu_n$ were calculated in~\cite{schiller2005,karr2006}. The next term corresponds to relativistic and QED corrections. The function $F_{\rm QED} (\alpha)$ is a nonanalytic expansion which, beyond powers of $\alpha$, also contains logarithmic terms like $\alpha^p \ln^q(\alpha)$. In principle, the coefficients of the expansion slightly depend on the mass ratios $\mu_n$, but this dependence may be neglected here. All coefficients have been calculated up to order $\alpha^3$ (or $R_{\infty} \alpha^5$ for the energy)~\cite{korobov2014a,korobov2014b}. The last term is the nuclear finite-size correction, which comprises a single term proportional to $r_p^2$ in H$_2^+$, and an additional term proportional to $r_d^2$ in HD$^+$~\cite{korobov2006}. The coefficients $A^{\rm fs}_n$ are proportional to the squared density of the wave function at the electron-nucleus coalescence point.

The dependence of a transition frequency $f$ on a fundamental constant $c$ is expressed in terms of a sensitivity coefficient
\begin{equation}
s_c = \frac{c_0}{f_0} \; \frac{\partial f}{\partial c} \label{defsensit},
\end{equation}
where $c_0$ is the recommended value of the fundamental constant $c$ and $f_0$ the transition frequency calculation for $c=c_0$ (and assuming recommended values for all other constants involved). All sensitivity coefficients of the selected transitions are given in Tables~\ref{sensith2} and~\ref{sensithd}. The sensitivities to $R_{\infty}$ (not shown) are very close to 1 and can be taken as equal to 1 for all practical purposes. The uncertainty due to $\alpha$ gives a negligible contribution to the overall uncertainty of the transition frequencies and will not be considered here.

The accuracy with which fundamental constants can be determined from the measurement of several rovibrational transitions depends on the uncertainty of those measurements, and of the related theoretical predictions from Eq.~(\ref{energy}). It is thus essential to give a realistic assessment of the accuracy that may be reached both in theory and experiments. Concerning theory, all correction terms of order $R_{\infty} \alpha^5$ have been calculated recently, leading to predictions of transition frequencies with $3-4 \times 10^{-11}$ relative uncertainty~\cite{korobov2014a,korobov2014b}). Based on current progress in the theoretical description, we estimate that the accuracy may be improved further by about one order of magnitude in the foreseeable future, and we will assume a theoretical uncertainty of $3 \times 10^{-12}$ for all transitions. This involves evaluating the following corrections: (i) two-loop self-energy at order $R_{\infty} \alpha^6$, (ii) nonlogarithmic one-loop self-energy correction of order $R_{\infty} \alpha^6$, and (iii) recoil corrections of order $R_{\infty} \alpha^4 (m/M)$, which are discussed in~\cite{jentschura2005,jentschura1999,pachucki1997} for the hydrogen atom case. Concerning the experimental accuracy on two-photon transition frequencies, we estimate that it may realistically reach a level of $1 \times 10^{-12}$~\cite{tran2013}. Indeed, the uncertainty associated with the various systematic frequency shifts (linear and quadratic Zeeman shifts, AC and DC Stark shifts, quadrupole shift, second-order Doppler shift) can be reduced well below this level (see e.g. the uncertainty budget in~\cite{biesheuvel2016}). All frequency measurements can be done with sufficient accuracy using optical frequency comb lasers~\cite{hansch2006,hall2006} referenced to commercially available cesium atomic clocks for traceability to the SI second.

We are now ready to estimate the uncertainty of $n$ fundamental constants $c_1 \ldots c_n$ extracted from the measurement of $n$ transition frequencies $f_1 \ldots f_n$. Here, we follow the approach of the CODATA least-squares adjustment (see the Appendices E and F in~\cite{codata1998}). Linearizing the expressions of the transition frequencies obtained from Eq.~(\ref{energy}) around the recommended values $c_{01} \ldots c_{0n}$ of the fundamental constants leads to the matrix relation
\begin{equation}
\mathbf{Y} = \mathbf{AX} \label{linear}
\end{equation}
where $\mathbf{X}$ and $\mathbf{Y}$ are column vectors with $n$ elements $x_1 \ldots x_n$ (resp. $y_1 \ldots y_n$) given by $x_j = (c_j - c_{0j})/c_{0j}$ (resp. $y_i = (f_i-f_{0i})/f_{0i}$, $f_{0i}$ being the frequency calculated for $c_j = c_{0j}$), and $A$ is a $n \times n$ matrix filled by the elements $a_{ij} = s^{i}_{c_j}$ (relative sensitivity of frequency $i$ on the constant $c_j$, as defined by Eq.~(\ref{defsensit})). Least-squares minimization gives us the best solution $\hat{\mathbf{X}}$ of Eq.~(\ref{linear}), for which the covariance matrix is~\cite{codata1998}
\begin{equation}
\mathbf{G} = \left( \mathbf{A}^T \mathbf{V}^{-1} \mathbf{A} \right)^{-1},
\end{equation}
and where $\mathbf{V}$ is the correlation matrix of the input data $\mathbf{Y}$. To construct this correlation matrix, we add the experimental and theoretical uncertainties quadratically. The experimental uncertainties of different transitions are assumed to be uncorrelated. However, theoretical uncertainties due to uncalculated terms are strongly correlated since these terms are primarily in the form of an unknown common constant multiplied by the square of the wave function at the electron-nucleus coalescence point~\cite{codata2010}. Here, we assume perfect correlations.

At this point, it is instructive to compare the relative uncertainties of individual transitions frequencies originating from each fundamental constant separately (correlations between constants are not considered for this evaluation). Taking the CODATA2014 uncertainties, one obtains for the $(0,2) \to (1,2)$ transition in H$_2^+$: $\left( \Delta y(R_{\infty}), \Delta y(\mu_{pe}), \Delta y(r_p) \right) = \left( 0.59, 4.4, 0.87 \right) \times 10^{-11}$. This confirms that $\mu_{pe}$, being the main source of uncertainty, is the parameter to be constrained from a measurement, as previously observed~\cite{wing1976,bressel2012,biesheuvel2016}. However, in the context of the proton-radius puzzle, it makes sense to set $\Delta r_p$ equal to the difference between the muonic hydrogen and CODATA values, and increase $\Delta R_{\infty}$ by the same factor as it is nearly perfectly correlated with $r_p$ (see the second line of Table~\ref{table-fc} for the values of the uncertainties). Then the contributions from the different constants are of the same order: $\left( \Delta y(R_{\infty}), \Delta y(\mu_{pe}), \Delta y(r_p) \right) = \left( 3.3, 4.4, 4.8 \right) \times 10^{-11}$, which shows that at least two other rovibrational transition frequency measurements are required to extract information on each constant separately. The situation is similar in HD$^+$; for the $(0,3) \to (9,3)$ transition one gets  $\left( \Delta y(R_{\infty}), \Delta y(\mu_{pe}), \Delta y(\mu_{dp}), \Delta y(r_p), \Delta y(r_d) \right) = \left( 0.59, 3.3, 1.1, 0.41, 0.42 \right) \times 10^{-11}$ with the CODATA uncertainties, and $\left( 3.3, 3.3, 1.1, 2.3, 2.6 \right) \times 10^{-11}$ when considering the discrepancies between nuclear radii.

\begin{table*}
\small
\begin{center}
\begin{tabular}{|c|c|c|c|c|c|c|}
\hline
                      & Used input & $10^{11} \; \Delta x(R_{\infty})$ &  $10^{11} \; \Delta x(\mu_{pe})$ & $10^{11} \; \Delta x(\mu_{dp})$ & $10^{3} \; \Delta x(r_p)$ & $10^{3} \; \Delta x(r_d)$ \\
\hline
                      & CODATA                            & 0.59 & 9.5  & 9.3  & 7.0 & 1.2 \\
                      & muonic atom discrepancy           & 3.3  &  -   &  -   & 39  & 6.3 \\
\hline
(i)                   & H2(0), H2(6), HD(R), HD(2), HD(9) & 0.86 & 0.82 & 5.6  & 8.4 & 3.3 \\
\hline
\multirow{2}{*}{(ii)} & H, H2(0), H2(6)                   & 1.8  & 1.6  &  -   & 22  &  -  \\
                      & H, H-D, H2(0), H2(6), HD(R)       & 1.8  & 1.6  & 3.4  & 22  & 3.7 \\
\hline
(iii)                 & $(r_p,r_d)$ + H2(0), HD(R), HD(9) & 0.34 & 0.41 & 0.84 & -   & -   \\
\hline
\multirow{2}{*}{(iv)} & $(R_{\infty},r_p)$ + H2(0)        &  -   & 0.68 &  -   & -   & -   \\
                      & $(R_{\infty},r_p,r_d)$ + H2(0), HD(9)  &  -   & 0.68 & 1.2  & -   & -   \\
\hline
\end{tabular}
\caption{Achievable relative accuracy on fundamental constants using HMI spectroscopy data (combined or not with atomic hydrogen or deuterium spectroscopy), assuming experimental and theoretical accuracies of $1 \times 10^{-12}$ and $3 \times 10^{-12}$ respectively. The first two lines refer to the present CODATA uncertainties and the discrepancies between electronic and muonic atom spectroscopy. The 'muonic' value of $r_d$ is obtained from the muonic hydrogen value of $r_p$, and using the determination of $r_d^2 - r_p^2$ from the 1S-2S isotopic shift measurement~\cite{jentschura2011}. Sections (i)-(iv) refer to different hypotheses on the outcome of the proton-radius puzzle (see text). \label{table-fc}}
\end{center}
\end{table*}

Our main results are summarized in Table~\ref{table-fc}, which is divided into four sections corresponding to different (hypothetical) outcomes of the proton-radius puzzle. In each case, we tested all possible combinations of transitions among those of Tables~\ref{sensith2} and~\ref{sensithd} and chose the one(s) leading to the most accurate determinations. The general guideline is to minimize redundancy, i.e. to select transitions having as diverse sensitivities as possible. For example, if two measurements in H$_2^+$ are required, the best choice is to combine the most different transitions in Table~\ref{sensith2}, which are $v=0 \to 1$ and $v=6 \to 7$.

We considered the following four (hypothetical) cases:

\noindent (i) {\em Puzzle unresolved: using only HMI data.} Five transition measurements in HMI yield a fully independent determination of $R_{\infty}$, $r_p$ and $r_d$. As can be seen by comparing the third line with the first two, the accuracy of $r_p, r_d$ and $R_{\infty}$ would approach that of the current CODATA values. Importantly, the results of HMI would provide enough resolution to shed light on the proton-radius puzzle as the uncertainties are significantly smaller than the related discrepancies. In addition, the uncertainty of $\mu_{pe}$ would be reduced by more than one order of magnitude over the present CODATA value, while the uncertainty of $\mu_{dp}$ would also be significantly improved.

\noindent (ii) {\em 1S-2S measurements in H and D confirmed: using 1S-2S and HMI data.} Combined with the H(1S-2S) result, two measurements in H$_2^+$ determine $R_{\infty}$, $\mu_{pe}$ and $r_p$. One additional measurement in HD$^+$, together with the H-D isotope shift measurement~\cite{jentschura2011}, allows determining also $\mu_{dp}$ and $r_d$. Again, the accuracy is good enough to shed light on the discrepancy, and the uncertainties of $\mu_{pe}$ and $\mu_{dp}$ are reduced by factors of 6 and 3, respectively.

\noindent (iii) {\em Muonic atom experiments confirmed: using muonic and HMI data.} Assuming that $r_p$ and $r_d$ are precisely determined by muonic atom spectroscopy, three HMI transition measurements allow determining $R_{\infty}$, $\mu_{pe}$ and $\mu_{dp}$. This would improve the uncertainty of both mass ratios by more than one order of magnitude, and that of the Rydberg constant by a factor of 1.7.

\noindent (iv) {\em Puzzle resolved: using 1S-2S, muonic and HMI data.} If muonic atom and hydrogen 1S-2S accuracies are confirmed, $r_p$, $r_d$ and $R_{\infty}$ are precisely determined independently of HMI data. We then revert to the initial idea of mass ratio determinations~\cite{wing1976,koelemeij2007}. A single measurement in H$_2^+$ improves $\mu_{pe}$ by a factor of 14, and an additional measurement in HD$^+$ yields a determination of $\mu_{dp}$ with an eightfold accuracy improvement.


We furthermore point out that an improved value of $\mu_{pe}$ may be combined with the accurate electron atomic mass determination reported by Sturm \textit{et al.}~\cite{sturm2014} to yield an improved value of the proton relative mass (reducing the uncertainty from $9\times 10^{-11}$ to $3 \times 10^{-11}$). In addition, combinations of accurate experimental and theoretical results of HMI spectroscopy can also be exploited to set greatly improved constraints on ''new physics'', such as the possible existence of a fifth force between hadrons~\cite{salumbides2013} or of compactified higher dimensions~\cite{salumbides2015}.

In conclusion, we have shown that Doppler-free two-photon spectroscopy of H$_2^+$ and HD$^+$ is a promising route to shed new light on the proton-radius puzzle. Depending on the progress and outcomes of ongoing experiments (atomic hydrogen spectroscopy, electron and muon scattering off nuclei), it may resolve the presently observed discrepancy, provide an alternative determination of the Rydberg constant, and improve the accuracy on the proton-electron and deuteron-proton mass ratios by one order of magnitude and beyond the 10$^{-11}$ level. We stress that the proposed approach is very attractive as it relies on Doppler-free frequency measurements of rovibrational transitions with extremely small natural linewidths, thus relaxing the requirement of a very precise understanding of the experimental lineshape. Similar to the role played by muonic hydrogen spectroscopy in the proton size puzzle, we expect that accurate theory and measurements of the HMI will provide essential input not only for the determination of fundamental constants, but also for foundational and cross-disciplinary checks of the validity of fundamental theory and experimental tests thereof, and for searches for new physics.

{\bf Acknowledgements.} We thank A. Antognini and F. Nez for helpful discussions. We gratefully acknowledge UPMC and the Russian Foundation for Basic Research under Grant No. 15-02-01906-a (V.I.K.), the Netherlands Technology Foundation (STW) and Foundation for Fundamental Research on Matter (FOM) (J.C.J.K.), the bilateral French-Dutch Van Gogh program, the Institut Universitaire de France (J.-Ph.K.), Agence Nationale de la Recherche under Grant ANR-13-IS04-0002-01 BESCOOL, and the European Commission under Programme FP7-PEOPLE-2013-ITN COMIQ (L.H.).

\end{document}